\documentclass[aps,prl,twocolumn,superscriptaddress,showpacs]{revtex4}

\usepackage{graphicx}

\usepackage{color}
\ifx\color@rgb\@undefined\else
  \definecolor{blue}{rgb}{0,0,1}
  \definecolor{red}{rgb}{1,0,0}
  \definecolor{comment}{rgb}{0,0.5,0}
  \definecolor{string}{rgb}{0,0.5,0.5}
\fi

\begin{document}


\title{
Investigation of the spin-glass regime between the antiferromagnetic and superconducting phases in
Fe$_{1+y}$Se$_x$Te$_{1-x}$}

\author{N. Katayama}
\affiliation{Department of Physics, University of Virginia, Charlottesville, Virginia 22904, USA}
\author{S. Ji}
\affiliation{Department of Physics, University of Virginia, Charlottesville, Virginia 22904, USA}
\author{D. Louca}
\affiliation{Department of Physics, University of Virginia, Charlottesville, Virginia 22904, USA}
\author{S.-H. Lee}
\affiliation{Department of Physics, University of Virginia, Charlottesville, Virginia 22904, USA}
\author{M. Fujita}					
\affiliation{Institute for Materials Research, Tohoku University, Sendai 980-8577, Japan}

\author{T. J. Sato}
\affiliation{The Institute for Solid State Physics, University of Tokyo, Kashiwa, Chiba 277-8581, Japan}

\author{J. S. Wen}
\affiliation{Condensed Matter Physics and Materials Science Department,
Brookhaven National Laboratory, Upton, New York 11973, USA}
\author{Z. J. Xu}
\affiliation{Condensed Matter Physics and Materials Science Department,
Brookhaven National Laboratory, Upton, New York 11973, USA}

\author{G. D. Gu}
\affiliation{Condensed Matter Physics and Materials Science Department,
Brookhaven National Laboratory, Upton, New York 11973, USA}
\author{G. Xu}
\affiliation{Condensed Matter Physics and Materials Science Department,
Brookhaven National Laboratory, Upton, New York 11973, USA}
\author{Z. W. Lin}
\affiliation{Condensed Matter Physics and Materials Science Department,
Brookhaven National Laboratory, Upton, New York 11973, USA}

\author{M. Enoki}					
\affiliation{Institute for Materials Research, Tohoku University, Sendai 980-8577, Japan}
\author{S. Chang}
\affiliation{NIST Center for Neutron Research, National Institute of Standards and Technology, Gaithersburg, Maryland 20899, USA}
\author{K. Yamada}
\affiliation{Institute for Materials Research, Tohoku University, Sendai 980-8577, Japan}
\author{J. M. Tranquada} 
\affiliation{Condensed Matter Physics and Materials Science Department,
Brookhaven National Laboratory, Upton, New York 11973, USA}

\date{\today}

\begin{abstract}
 Using bulk magnetization along with elastic and inelastic neutron scattering techniques, we have investigated the phase diagram of Fe$_{1+y}$Se$_{x}$Te$_{1-x}$ and the nature of magnetic correlations in three nonsuperconducting samples of Fe$_{1.01}$Se$_{0.1}$Te$_{0.9}$, Fe$_{1.01}$Se$_{0.15}$Te$_{0.85}$ and Fe$_{1.02}$Se$_{0.3}$Te$_{0.7}$.  A cusp and hysteresis in the temperature dependence of the magnetization for the $x=0.15$ and 0.3 samples indicates spin-glass (SG) ordering below $T_{\rm sg} = 23$~K.  Neutron scattering measurements indicate that the spin-glass behavior is associated with short-range spin density wave (SDW) ordering characterized by a static component and a low-energy dynamic component with a characteristic incommensurate wave vector of ${\bf Q}_m = (0.46, 0, 0.50)$ and an anisotropy gap of $\sim$ 2.5 meV. Our high ${\bf Q}$-resolution data also show that the systems undergo a glassy structural distortion that coincides with the short-range SDW order.
 \end{abstract}

\pacs{71.30.+h, 61.14.-x, 61.50.Ks, 61.66.Fn}
\maketitle
Following the discovery of superconductivity in Fe-based pnictides,\cite{rf:discovery} a resurgence of interest in the field of high temperature superconductivity ensued.\cite{rf:Ren, rf:Chen, rf:Sachdev, rf:Mazin, rf:Chubukov, rf:ZT, rf:Cruz} 
There has been particular interest in the possible connection between magnetism and superconductivity.  In the iron pnictides, an antiferromagnetically ordered phase is in close proximity to optimal superconductivity.\cite{rf:Ishida}  In some cases, such as SmFeAsO$_{1-x}$F$_x$ and Ba(Fe$_{1-x}$Co$_x$)$_2$As$_2$, there is evidence for coexisting antiferromagnetic order and superconductivity.\cite{rf:Drew,rf:Bernhard,rf:Chu,rf:Laplace,rf:Fernandez}  The situation is somewhat different in the chalcogenide system, Fe$_{1+y}$Se$_x$Te$_{1-x}$.  Here the details are sensitive to the Fe as well as the Se concentration, and we will focus on the situation for minimized excess Fe ({\it i.e.}, $y\approx 0$).  The N\'eel temperature drops rapidly for $x\alt 0.1$, but our measurements indicate that bulk superconductivity only appears for $x\agt0.4$.  

One reason for a difference between the pnictides and chalcogenides concerns the nature of the antiferromagnetic order.  To discuss that order, we first have to consider the crystal structure.  In the $\alpha$-PbO structure of Fe$_{1+y}$Se$_x$Te$_{1-x}$ (FST), the Fe layers have a square lattice structure; however, the positions of the Se/Te atoms above and below those planes break the translational symmetry.  Thus, it is crystallographically appropriate to choose a unit cell with two Fe atoms per layer, such that the lattice parameter is $a\approx 3.8$~\AA.  We will specify reciprocal lattice vectors, ${\bf Q}=(h,k,l)$, in reciprocal lattice units (rlu) of $(2\pi/a,2\pi/b,2\pi/c)$.  In Fe$_{1+y}$Te, the long-range SDW state is accompanied by a tetragonal-to-monoclinic (or orthorhombic, depending on $y$) structural transition.\cite{rf:Bao, rf:Li}  The spin arrangement is ferromagnetic along the $b$-direction and alternates in a  $++--$ fashion along the $a$-direction, leading to a characteristic wave vector of (0.5,0,0.5).  For larger $y$ ({\it e.g.}, $y=0.14$), the in-plane component of the magnetic wave vector becomes slightly incommensurate.\cite{rf:Bao} In Fe$_{1+y}$Se$_{x}$Te$_{1-x}$ with $0.25\le x\le0.33$,  static, but short-range, incommensurate magnetic order with ${\bf Q}_m = (0.5-\delta, 0, 0.5)$ is observed.\cite{rf:Bao,rf:Jinsheng,rf:Khasanov}  At higher Se concentration, $x \agt 0.4$, where bulk superconductivity is achieved, a spin resonance at $\hbar\omega \simeq$ 6.5 meV appears at incommensurate ${\bf Q}_c = (0.5\pm \delta', 0.5\mp \delta',l)$.\cite{rf:Qiu,rf:Lumsden,rf:Argyriou,rf:Seunghun, rf:Dai}

The reduction in crystallographic symmetry is important for the magnetic ordering in Fe$_{1+y}$Te.  The monoclinic (or orthorhombic) structure provides the magnetic ordering wave vector with a unique orientation within the Fe planes.  On the other hand, the short-range magnetic order observed at $x\sim0.3$ occurs in a tetragonal phase, so that there are two degenerate orientations for ${\bf Q}_m$.  This suggests that competition among degenerate domains may lead to frustration and keep the ordering short range.  A recent study\cite{rf:Martinelli} has shown that the crossover to the tetragonal phase occurs between $x=0.075$ and $x=0.10$, with long-range magnetic order only for $x\le0.075$. Thus, one might expect a transition to short-range SDW order for $x\agt0.1$.

In this paper, we show that spin-glass-like behavior is present in FST for $x=0.1$ and 0.15.  We present evidence from magnetization measurements and characterize the short-range order with neutron scattering.  One of our main results is that the short-range order is structural as well as magnetic, consistent with the proposal that orbital ordering is an important part of the magnetically-ordered state \cite{rf:yin}. We have also studied the low-energy spin fluctuations for $x = 0.15$ and 0.3 with inelastic neutron scattering.  While there is some weak critical scattering that extends out to ${\bf Q}_c$ near the onset of elastic magnetic scattering, that disappears at low temperature, as the spin fluctuations are dominantly associated with ${\bf Q}_m$.  Thus, there appears to be a broad spin-glass (SG) regime in FST associated with a type of geometrical frustration.  The eventual onset of bulk superconductivity appears to be associated with an evolution of the characteristic wave vector associated with the spin fluctuations.

\begin{figure}
\center
\includegraphics[width=8.0cm]{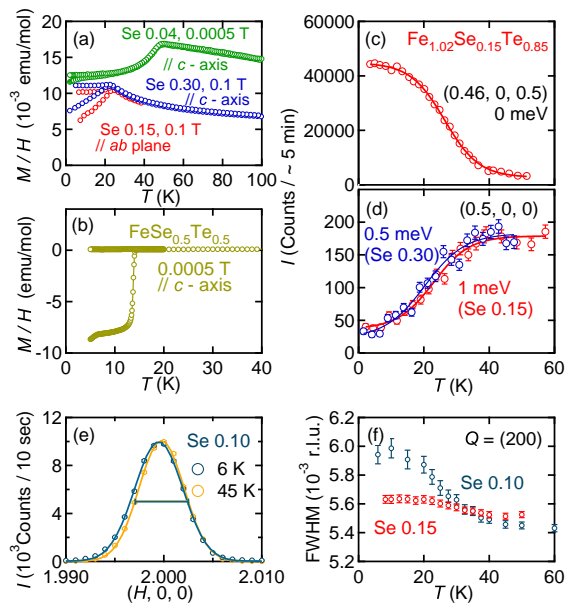}
\caption{\label{fig:Fig1}
(Color online) (a), (b) Bulk magnetic susceptibility data of Fe$_{1.08}$Se$_{0.04}$Te$_{0.96}$ (green), Fe$_{1.02}$Se$_{0.15}$Te$_{0.85}$ (red), Fe$_{1.01}$Se$_{0.3}$Te$_{0.7}$ (blue) and FeSe$_{0.5}$Te$_{0.5}$ (light green circles), measured under the external magnetic field of 5 Gauss along the $c$-axis, 100 Gauss along the $c$-axis, 100 Gauss perpendicular to the $c$-axis, and 5 Gauss along the $c$-axis, respectively. (c), (d) Neutron measurements data for $x = 0.15$ (red) and $x = 0.3$ (blue). (c) Temperature ($T$) dependence of the elastic neutron scattering intensity at ${\bf Q}_m = (0.46,0,0.5)$ measured for $x = 0.15$. (d) $T$-dependence of the inelastic neutron scattering intensities measured with $\hbar\omega =$ 1 meV for $x = 0.15$ (red circles) and with $\hbar\omega = 0.5$ meV for $x=0.3$ (blue circles). (e) Elastic longitudinal scans at ${\bf Q} =$ (2,0,0) for Fe$_{1.01}$Se$_{0.1}$Te$_{0.9}$ measured at 6 K and 45 K. The lines are fit to a Gaussian function. The horizontal bar represents the instrumental ${\bf Q}$-resolution. (f) Full-Width-of-the-Half-Maximum (FWHM) vs temperature obtained for $x=0.1$ and $x=0.15$.}
\end{figure}

Single crystals of FST with various Se concentrations were
prepared using an unidirectional solidification method 
at Brookhaven National Laboratory.  In addition, crystals of Fe$_{1.01}$Se$_{0.1}$Te$_{0.9}$ and Fe$_{1.01}$Se$_{0.3}$Te$_{0.7}$ were grown at the Institute for Solid State Physics, University of Tokyo.
For bulk magnetization measurements, $\sim$ 0.01 g single crystals with various Se concentrations from $x = 0$ to $x = 0.7$ were used in a superconducting quantum interference device (SQUID) magnetometer, while for neutron measurements, a 0.39 g single crystal of  Fe$_{1.01}$Se$_{0.1}$Te$_{0.9}$, a 10.1 g single crystal of FeSe$_{0.15}$Te$_{0.85}$ and a 5.3 g single crystal of Fe$_{1.01}$Se$_{0.3}$Te$_{0.7}$ were used. The elemental concentrations for the crystals used for neutron scattering were determined by energy dispersive X-ray spectroscopy, while nominal concentrations, based on ratios of starting materials, are used for the other crystals. 

The neutron scattering experiments were performed at the cold-neutron triple-axis spectrometer SPINS, at the NIST Center for Neutron Research.  Most of the experiments on the $x=0.15$ and 0.3 single crystals were done with the instrumental configuration of guide--open--$80'$--open and the energy of the scattered neutrons fixed to $E_f=$ 5 meV. One Be filter cooled by liquid nitrogen was placed after the sample to minimize higher order neutron contamination. An additional Be filter was placed in front of the sample for the elastic measurements. The $x=0.15$ single crystal was aligned in the $(h,k,0)$ and the $(h,0,l)$ planes, while the $x=0.3$ single crystal was aligned in the $(h,k,0)$ plane. High ${\bf Q}$-resolution elastic measurements on the $x=0.1$ and $x=0.15$ single crystals were performed using a backscattering geometry with the instrumental configuration of guide--$20'$--$20'$--$40'$ and $E_i = 10$ meV.

Figure \ref{fig:Fig1} (a) and (b) show the bulk magnetic susceptibility data obtained from single crystals of FST with four different Se concentrations, $x = 0.04, 0.15, 0.3$ and 0.5. For $x=0.04$, a sharp decrease is observed at $T_{\rm SDW} \simeq 49$ K, indicating a long-range magnetic order as reported in the pure Fe$_{1+y}$Te compound.  For $x=0.15$ and 0.3, on the other hand, the sharp decrease is replaced by a cusp at $T_{\rm sg} \sim  23$ K, accompanied by an FC-ZFC hysteresis below, indicating that the magnetic ordering is short range.  In the spin-glassy compounds, the transition is second-order, with neither observable long-range structural phase transition nor superconducting transition. Upon further doping of Se ions, the system becomes superconducting as shown in Fig. 1 (b) for $x=0.5$ with the superconducting phase transition temperature of $T_c = 14.5$ K.

\begin{figure}
\center
\includegraphics[width=8.0cm]{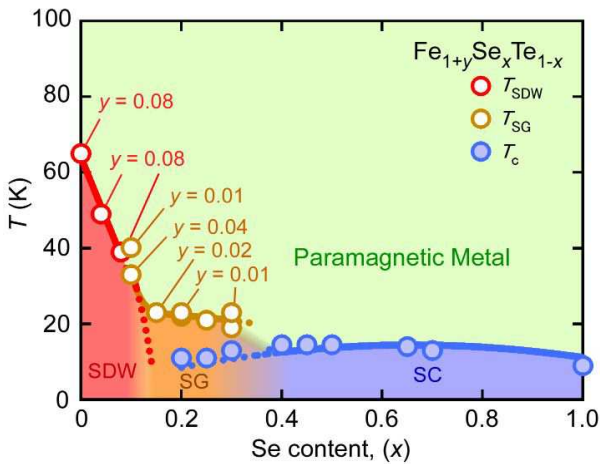}
\caption{\label{fig:Fig4-4}
(Color online)  Phase diagram of Fe$_{1+y}$Se$_x$Te$_{1-x}$ with $y \sim 0$ as a function of $x$ and $T$, constructed from single crystal bulk susceptibility data some of which are shown in Fig. 1 (a) and (b), except for $x =$ 1 which is taken from Refs. \cite{rf:Hsu, rf:McQueen}. The nominal Fe content, $y$, is $y = 0$ unless it is specified. $T_c$ (blue circles) represents the superconducting onset temperature.}
\end{figure}

Figure 2 shows the $x$--$T$ phase diagram for FST based only on the bulk susceptibility data obtained from the single crystal samples. Even though the values of $x$ and $y$ are nominal values and may not be exactly correct, the phase diagram clearly shows the trends and the existence of three distinct phases; the antiferromagnetic phase for $x \alt 0.1$, the bulk superconducting phase for $x \agt 0.4$, and the intermediate spin-glass phase. Our phase diagram clearly shows that the long-range ordered SDW phase is non-superconducting. While the original paper by Fang {\it et al.}\cite{rf:Fang} reported superconductivity in the same phase based on powder samples, problems with contamination by oxide phases in that work have already been pointed out by McQueen {\it et al}.\cite{rf:McQueen}  In the intermediate phase, some samples showed partial superconductivity below $T_c \sim 11$ K as shown the figure, while  others were non-superconducting down to 1.4 K. In this paper, we focus on the magnetic character of these SG samples.

Firstly, in order to investigate what happens to the crystal structure at low temperatures in the spin-glassy phase, we have performed high ${\bf Q}$-resolution elastic measurements on the $x=0.1$ and 0.15 single crystals. Fig. 1 (e) shows the results of the longitudinal scans obtained for $x=0.1$ at the nuclear (2,0,0) Bragg reflection. The peak does not split into two peaks at 6 K, but the 6 K peak is clearly broader than that of 45 K.  The low-temperature broadening suggests a structural tendency towards lower symmetry.
The same scans were done at several different temperatures, and the data were fit to a single Gaussian. The Full-Width-of-the-Half-Maximum (FWHM) is plotted as a function of $T$ for both Se concentrations in Fig. 1 (f). The structural modification develops below 40 K and weakens with increasing Se concentration.

\begin{figure}
\center
\includegraphics[width=8.cm]{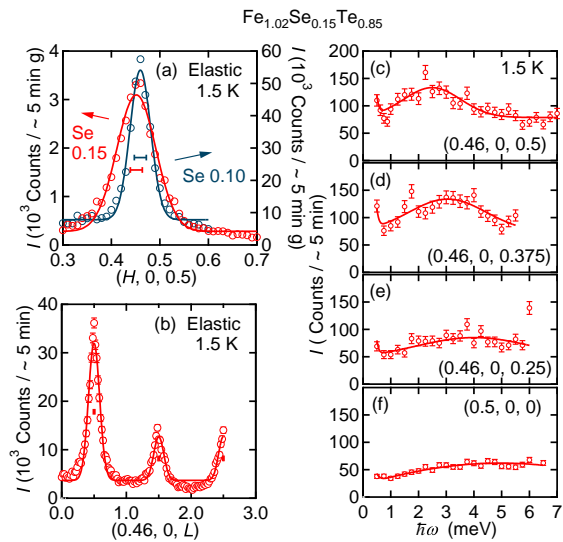}
\caption{\label{fig:Fig2-4}
(Color online) Elastic neutron scattering data obtained at 1.5 K for $x=0.15$ (a) along $(H,0,0.5)$ and (b) $(0.46,0,L)$ directions. (c)-(f) Inelastic neutron scattering data as a function of $\hbar\omega$ at various different wave vectors along the $L$-direction. In (a), data of $x=0.1$ is also shown. Error bars for the neutron data represent one standard deviation.}\end{figure}

Secondly, the static spin correlations in the spin-glass phase were investigated using elastic neutron scattering.  As shown in Fig.~\ref{fig:Fig2-4} (a) and (b), scans on $x=0.15$ at 1.5~K reveal prominent static magnetic peaks at incommensurate wave vectors ${\bf Q} = (0.46,0,L+0.5)$ for integer $L$. On cooling from higher temperatures, the static spin correlations gradually freeze below $T_f = 40$~K [Fig.~\ref{fig:Fig1} (c)]. The increase of the static spin correlations coincides with the reduction in the intensity of the low energy excitations [Fig. \ref{fig:Fig1} (d)]. The $T_f$ measured by elastic neutron scattering with an energy resolution of $\Delta E \sim$ 0.3 meV is higher than the $T_{\rm sg}$ measured by static bulk susceptibility, which is common in systems involving spin freezing.\cite{rf:Chou,rf:Wakimoto}  
At 1.5 K, the peak width is 0.103(6) r.l.u. along [100] and 0.209(4) r.l.u. along [001], which are larger than the instrumental resolution.  The correlation length is estimated to be $\xi_a =$ 12(1) {\AA} along [100] and 9.5(2) {\AA} along [001], which is consistent with the reported data of $x=0.25$.\cite{rf:Jinsheng} It is to be noted that the magnetic ordering of $x=0.15$ is more short-ranged and weaker than that of $x=0.1$ with $\xi_a=$ 24(6) {\AA} (Fig. 2 (a)). 

\begin{figure}
\center
\includegraphics[width=8.cm]{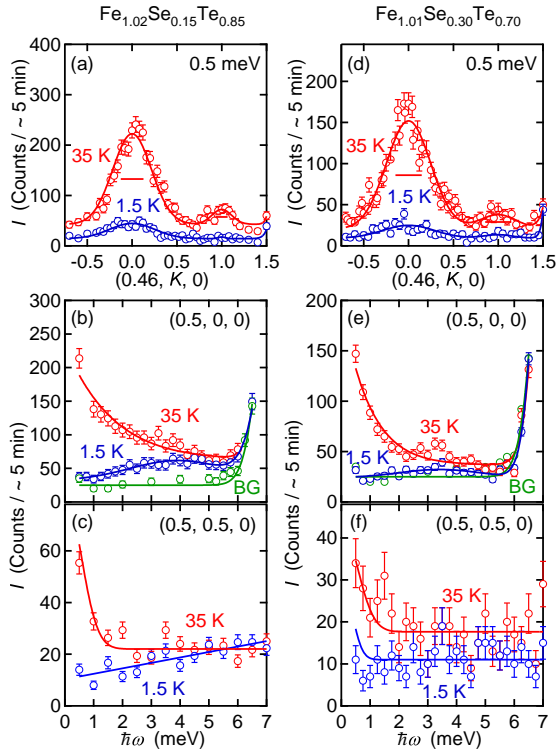}
\caption{\label{fig:Fig3-4}
(Color online)  Inelastic neutron scattering data obtained (a)-(c) for
$x=0.15$ and (d)-(f) for $x=0.3$. (a), (d) show the constant-$\hbar\omega = 0.5$ meV scan along the $(0.46,K)$ direction. (b), (e) show the constant-${\bf Q} = (0.5,0)$ scan as a function of $\hbar\omega$, while (c), (f) show the constant-${\bf Q} = (0.5,0.5)$ scan. The measurements were done at 35 K (red circles) and 1.5 K (blue circles). Green circles in (b) and (c) are the background measured with the crystal rotated by 30 degrees. The rapid increase in the background in (b) is due to the fact that the scattering angle decreases as $\hbar\omega$ increases, so that the tail of the direct neutron beam starts to contribute at high energies for fixed {\bf Q}.}
\end{figure}

Let us now turn to the dynamic spin correlations. The constant-{\bf Q} scans as a function of $\hbar\omega$ at various $l$ values, shown in Fig. 2 (c)-(f), indicate that there is an anisotropy gap $\Delta \sim 2.5$~meV. As $l$ is decreased from 0.5, $\Delta$ disperses to higher energies, reaching $\sim 4$ meV at $l=0$. The weak dispersion of $\Delta$ along [001] indicates that the magnetic interactions between the Fe-layers along the $c$-axis are weak, resulting in quasi two-dimensional (2D) behavior.

In order to test for evidence of low energy magnetic excitations near ${\bf Q}_c$, characteristic of the superconducting phase at larger $x$, we reoriented the crystals into the $(hk0)$ scattering plane and performed further inelastic scattering measurements, taking advantage of the quasi-2D character of the magnetic correlations.
Figure~3(a) and (d) show the presence of spin fluctuations along ${\bf Q}=(0.46, k,0)$ at an energy transfer of $\hbar$$\omega$ = 0.5 meV for $x=0.15$ and $x=0.3$, respectively. At 35 K, which is above $T_{\rm sg}$ but below $T_f$,  the inelastic signal peaks at $k=0$ and 1, for both compositions.  Note that the magnetic scattering has a minimum at $k=0.5$, corresponding to ${\bf Q}_c$, in contrast to the behavior of superconducting samples.
At 1.5 K, the peak intensity is suppressed due to weight transfer to the elastic channel.  

The energy dependence of the spin fluctuations measured at ${\bf Q} = (0.5,0,0)$ is shown in Fig.~3 (b) and (e).  The  energy scans up to 7 meV show a strong quasielastic signal at 35 K.  By 1.5 K, the low energy spectral weight is depleted, while a broad peak remains at $\sim 4$ meV, corresponding to $\Delta$ as shown in Fig.~2(f).  Figure 3(c) and (f) show no obvious excitation at ${\bf Q} = (0.5, 0.5,0)$ up to 7 meV.  The weak enhancement observed at 35 K is the tail of the strong spin fluctuations centered at (0.5,0,0) as shown in Fig.~3(a) and (d). Thus, we conclude that there is no significant low-energy magnetic response at ${\bf Q}_c$ in nonsuperconducting FST with $x\alt 0.3$.

To summarize, we presented  the $x$--$T$ phase diagram of Fe$_{1+y}$Se$_x$Te$_{1-x}$ based on our systematic bulk susceptibility measurements on single crystals. We also performed elastic and inelastic neutron scattering measurements on non-superconducting single crystals of Fe$_{1.01}$Se$_{0.1}$Te$_{0.9}$, Fe$_{1.02}$Se$_{0.15}$Te$_{0.85}$ and Fe$_{1.01}$Se$_{0.30}$Te$_{0.70}$. All samples exhibit a static short-range incommensurate SDW transition with the characteristic wave vector ${\bf Q}_m = (0.46, 0, 0.5)$. Our high ${\bf Q}$-resolution data on the (200) nuclear Bragg reflection show that the development of the short range SDW ordering coincides with that of a crystal structural distortion presumably involving reduced symmetry. The low energy magnetic excitations below and above $T_{\rm sg}$ can be characterized by excitations centered around ${\bf Q}_m$. Even for Fe$_{1.01}$Se$_{0.30}$Te$_{0.70}$, which is close to the SC phase of Fe$_{1+y}$Se$_x$Te$_{1-x}$, no low-energy dynamic spin correlations were present around ${\bf Q}_c = (0.5,0.5,0)$. 

Work at the University of Virginia (Brookhaven) is supported by the U.S. Department of Energy, Office of Basic Energy Sciences, Division of Materials Sciences and Engineering, under Contract Nos. DE-FG02-07ER45384 and DE-FG02-01ER45927 (DE-AC02-98CH110886).  This work utilized facilities supported in part by the National Science Foundation under Agreement No. DMR-0454672.

\end{document}